\documentclass[aps,preprint]{revtex4}%
\usepackage{amsfonts}
\usepackage{amsmath}
\usepackage{amssymb}
\usepackage{graphicx}%
\providecommand{\U}[1]{\protect\rule{.1in}{.1in}}

\begin{document}
\title[ ]{Contrasting Interactions Between Dipole Oscillators in Classical and Quantum
Theories: Illustrations of Unretarded van der Waals Forces}
\author{Timothy H. Boyer}
\affiliation{Department of Physics, City College of the City University of New York, New
York, New York 10031}

\begin{abstract}
Students encounter harmonic-oscillator models in many aspects of basic
physics, within widely-varying theoretical contexts. \ Here we highlight the
interconnections and varying points of view. We start with the classical
mechanics of masses coupled by springs and trace how the same essential
systems are reanalyzed in the unretarded van der Waals interactions between
dipole oscillators within classical and quantum theories. \ We note how
classical mechanical ideas from kinetic theory lead to energy equipartition
which determines the high-temperature van der Waals forces of atoms and
molecules modeled as dipole oscillators. \ In this case, colliding heat-bath
particles can be regarded as providing \textit{local} hidden variables for the
statistical mechanical behavior of the oscillators. \ Next we note how
\textit{relativistic} classical electrodynamical ideas conflict with the
assumptions of \textit{nonrelativistic} classical statistical mechanics.
\ Classical electrodynamics which includes classical zero-point radiation
leads to van der Waals forces between dipole oscillators, and these classical
forces agree at all temperatures with the forces derived from quantum theory.
\ However, the classical theory providing this agreement is not a
\textit{local} theory, but rather a \textit{non-local} hidden-variables
theory. \ The classical theory can be regarded as involving hidden variables
in the random phases of the plane waves spreading throughout space which
provide the source-free random radiation.

\end{abstract}
\maketitle

\section{Introduction}

Students see harmonic-oscillator systems in many of their physics classes.
\ Individual oscillators and coupled oscillators appear in classical
mechanics, in electrodynamics, in statistical mechanics, in quantum mechanics,
and in modern physics. \ The oscillator Hamiltonians involved in these areas
may be very similar or even identical, but the theoretical contexts behind the
oscillator systems can be strikingly different. \ Charged harmonic oscillators
are often used as simple models of atoms and molecules, and the classical and
quantum interpretations vary sharply. \ Here we illustrate some of the
contrasting ideas of classical and quantum physics, with the unretarded van
der Waals forces between dipole oscillator systems at finite temperature
providing the unifying element. \ We believe that both students and
instructors will be interested in the contrasts in the theoretical contexts
which often go unmentioned. \ We start by recalling aspects of harmonic
oscillators in classical mechanics which every advanced undergraduate student
has seen. \ We then turn to oscillators in thermal equilibrium within three
different theoretical contexts: classical statistical mechanics, classical
electrodynamics, and quantum theory. \ 

\section{Interacting Dipole Oscillators In Classical Mechanics}

\subsection{Single Oscillator}

The description of our interacting dipole oscillators begins where students
begin, namely with classical mechanics. \ The Hamiltonian for a single
harmonic oscillator of mass $m$ and natural (angular) frequency $\omega_{0}$
with displacement $x$ and momentum $p,$ is given by
\begin{equation}
H=\frac{p^{2}}{2m}+\frac{1}{2}m\omega_{0}^{2}x^{2}. \label{Ham1}%
\end{equation}
The Hamiltonian equations of motion $\dot{x}=\partial H/\partial p=p/m$ and
$\dot{p}=-\partial H/\partial x=-m\omega_{0}^{2}x\,$yield the harmonic
oscillator solution
\begin{equation}
x(t)=\sqrt{\frac{2\mathcal{E}}{m\omega_{0}^{2}}}\cos(\omega_{0}t+\phi)
\label{xosc}%
\end{equation}
where the constant $\mathcal{E}$ corresponds to the energy of the oscillator
\begin{equation}
\mathcal{E=}\frac{1}{2}m\dot{x}^{2}+\frac{1}{2}m\omega_{0}^{2}x^{2}%
\end{equation}
and the constant $\phi$ is an arbitrary phase. \ The time average of the
kinetic energy equals the time average of the potential energy,%
\begin{align}
\left\langle \frac{1}{2}m\dot{x}^{2}\right\rangle  &  =\frac{1}{2}m\omega
_{0}^{2}\frac{2\mathcal{E}}{m\omega_{0}^{2}}\left\langle \sin^{2}(\omega
_{0}t+\phi)\right\rangle =\frac{1}{2}\mathcal{E},\nonumber\\
\left\langle \frac{1}{2}m\omega_{0}^{2}x^{2}\right\rangle  &  =\frac{1}%
{2}m\omega_{0}^{2}\frac{2\mathcal{E}}{m\omega_{0}^{2}}\left\langle \cos
^{2}(\omega_{0}t+\phi)\right\rangle =\frac{1}{2}\mathcal{E}, \label{Eav}%
\end{align}
since the average of the squares of the sine and cosine functions are both
$1/2.$

\subsection{Two Interacting Dipole Oscillators}

\subsubsection{Interacting Hamiltonian}

Next we consider two interacting dipole oscillators, since this provides the
unifying model for our discussion. \ Interacting harmonic oscillators are
often used in the treatments of unretarded van der Waals forces between atoms
and molecules. \ We introduce an electric dipole interaction\cite{Griffiths1}
$-2e^{2}x_{A}x_{B}/R^{3}$ between two identical charged harmonic oscillators
$A$ and $B$ separated by a distance $R$ along the $x$-axis which is parallel
to the oscillation direction for each oscillator, corresponding to the
situation in a standard quantum mechanics text.\cite{GriffithsQ} \ For this
situation, the Hamiltonian takes the form\cite{GriffithsQ}%
\begin{equation}
H=\frac{p_{A}^{2}}{2m}+\frac{1}{2}m\omega_{0}^{2}x_{A}^{2}+\frac{p_{B}^{2}%
}{2m}+\frac{1}{2}m\omega_{0}^{2}x_{B}^{2}-\frac{2e^{2}x_{A}x_{B}}{R^{3}%
}.\label{Hint1}%
\end{equation}
Since we are still in the classical mechanics section of our discussion, we
note that this same Hamiltonian can be interpreted in terms of masses
interacting through springs. \ In the classical mechanics texts, this
Hamiltonian corresponds to two particles of equal mass $m$, one of which is
attached to a wall to the left and the other is attached to a wall to the
right by springs of spring constant $\kappa,$ while the masses are coupled
together by a spring between the masses of spring constant $\kappa^{\prime},$
where $\kappa=m\omega_{0}^{2}-2e^{2}/R^{3}$ and $\kappa^{\prime}=2e^{2}/R^{3}%
$.\cite{Fowles} \ Proceeding with the usual classical mechanical treatment,
the Hamiltonian can be rewritten in terms of the normal coordinates $x_{+}$
and $x_{-}$ given by
\begin{equation}
x_{\pm}=\frac{1}{\sqrt{2}}\left(  x_{A}\pm x_{B}\right)  \text{ \ \ \ and
\ \ \ \ }p_{\pm}=\frac{1}{\sqrt{2}}\left(  p_{A}\pm p_{B}\right)  ,\text{
}\label{norm}%
\end{equation}
so that the Hamiltonian now takes the form\cite{GriffithsQ}
\begin{equation}
H=\left[  \frac{p_{+}^{2}}{2m}+\frac{1}{2}\left(  m\omega_{0}^{2}-\frac
{2e^{2}}{R^{3}}\right)  x_{+}^{2}\right]  +\left[  \frac{p_{-}^{2}}{2m}%
+\frac{1}{2}\left(  m\omega_{0}^{2}+\frac{2e^{2}}{R^{3}}\right)  x_{-}%
^{2}\right]  ,\label{Hint2}%
\end{equation}
corresponding to two uncoupled oscillators with associated (angular)
frequencies of oscillation given by%
\begin{equation}
\omega_{\pm}^{2}=\omega_{0}^{2}\mp\frac{2e^{2}}{mR^{3}}.\label{omegpm}%
\end{equation}
Note the arrangement of the $\pm$ signs. \ In the symmetric normal mode
labeled by $x_{+},$ the oscillators move in-phase. \ Thus in the
classical-mechanical spring model, the central coupling spring is not
compressed in this mode; in the electrostatic force model, the
inter-oscillator electric dipole field tends to oppose the spring restoring
force on each oscillator. \ Hence the interaction frequency $\omega_{+}$ is
the lower frequency. \ The situation is reversed for the antisymmetric $x_{-}$
normal mode, and the oscillation frequency $\omega_{-}$ has the higher frequency.

The electric dipole oscillators in the Hamiltonian of Eq. (\ref{Hint1}) are
treated in the electrostatic approximation which includes the electrostatic
interaction between the dipole oscillators but which neglects any radiation
emission which would carry energy away from the interacting dipole oscillators.

\subsubsection{Normal Modes of Oscillation}

The normal mode solutions can be written in the form given above in Eq.
(\ref{xosc}) for a single oscillator
\begin{equation}
x_{+}=\sqrt{\frac{2\mathcal{E}_{+}}{m\omega_{+}^{2}}}\cos(\omega_{+}t+\phi
_{+}),\text{ \ \ }x_{-}=\sqrt{\frac{2\mathcal{E}_{-}}{m\omega_{-}^{2}}}%
\cos(\omega_{-}t+\phi_{-}). \label{xpxm}%
\end{equation}
where $\mathcal{E}_{+}$ and $\mathcal{E}_{-}$ are the energies associated with
the normal modes of oscillation of the interacting system.

Then writing $x_{A}=(x_{+}+x_{-})/\sqrt{2},~~x_{B}=(x_{+}-x_{-})/\sqrt{2}$ and
using the expressions in Eq. (\ref{xpxm}), we have the solutions for the
motions of the two oscillators $A$ and $B.$ \ The two oscillators exchange
energy\cite{Fowles} through the interaction term in Eq. (\ref{Hint1}). \ If we
start with all the energy in one oscillator and none in the second, the
displacement of the first oscillator provides a force on the second, so that,
after a suitable time interval, we will find that all the energy has been
transferred to the second oscillator and there is none in the
first.\cite{Fowles} \ The two identical oscillators exchange energy, with the
(angular) frequency of exchange given by the beat frequency $\omega
_{exchange}=\omega_{+}-\omega_{-},$ which, from Eq. (\ref{omegpm}), depends on
the strength of the electric dipole interaction $-(2e^{2}/R^{3})x_{A}x_{B},$
corresponding to the strength of the spring constant $\kappa^{\prime}$ in the
classical mechanical texts.\cite{Fowles} \ As the interaction between the
dipole oscillators becomes weaker, the energy exchange frequency
$\omega_{exchange}$ decreases, vanishing in the uncoupled limit. \ However,
for any finite coupling, no matter how small, the energy is always exchanged
between the two oscillators $A$ and $B$. \ 

\subsubsection{Correlation Between the Oscillators}

We will be interested in the correlations between displacements of the
spatially-separated oscillators. \ Thus for our two interacting oscillators,
we can calculate the time average of the product of the oscillator
displacements using Eq. (\ref{xpxm}). \ We find
\begin{equation}
\left\langle x_{A}x_{B}\right\rangle =\left\langle (x_{+}+x_{-})(x_{+}%
-x_{-})/2\right\rangle =\left\langle x_{+}^{2}-x_{-}^{2}\right\rangle
/2=\frac{\mathcal{E}_{+}}{2m\omega_{+}^{2}}-\frac{\mathcal{E}_{-}}%
{2m\omega_{-}^{2}}. \label{xAxBcl}%
\end{equation}
Evidently, the spatial correlation function for the two oscillator
displacements depends upon both the energies $\mathcal{E}_{+},$ $\mathcal{E}%
_{-}$ in the normal modes and also the frequencies $\omega_{+},$ $\omega_{-}$
of the normal modes. \ For example, if all the energy is in the symmetric mode
with frequency $\omega_{+}$ and none in the antisymmetric mode $\omega_{-},$
then the two oscillators move together in phase, and, in this case,
$\left\langle x_{A}x_{B}\right\rangle =\mathcal{E}_{+}/(2m\omega_{+}^{2})$ is
positive. \ The magnitude of the correlation depends upon the amplitude of the
$x_{+}$-motion given in Eq. (\ref{xpxm}) which involves $[\mathcal{E}%
_{+}/(m\omega_{+}^{2})]^{1/2}.$ \ The sign of the correlation is reversed for
the antisymmetric mode. \ Thus the correlation in Eq. (\ref{xAxBcl}) has an
immediate natural interpretation in terms of the normal modes of oscillation
of the system.

\section{Interacting Dipole Oscillators in Classical Statistical Mechanics}

Having reviewed the classical mechanical behavior of harmonic oscillators, we
now reexamine the oscillators' behavior in the theoretical context of
classical statistical mechanics. \ 

\subsection{Harmonic Oscillators as Models for van der Waals Forces}

The harmonic oscillator systems, which involve masses coupled by springs in
the classical mechanical texts, reappear in modern physics texts\cite{Eisberg}
in connection with van der Waals forces between atoms and molecules which are
modeled as fluctuating electric dipole oscillators. \ The oscillations of the
dipole-oscillator models are taken as those associated with thermal
equilibrium. \ The forces between the dipole oscillators are regarded as
modeling the forces between neutral atoms and molecules.

\subsection{Single Oscillator in a Heat Bath}

The basic classical mechanical ideas for dipole oscillators, which we
considered above, reappear in classical statistical mechanics. We emphasize
that the \textquotedblleft mechanics\textquotedblright\ both of the
oscillators and of classical statistical mechanics are part of
\textit{nonrelativistic theory, }and that classical statistical mechanics
might well be termed \textquotedblleft\textit{nonrelativistic} classical
statistical mechanics.\textquotedblright\ \ 

\subsubsection{Understanding Energy Equipartition}

Classical statistical mechanics was developed during the 19th century by the
application of statistical ideas to classical mechanics . \ The theory is an
outgrowth of classical kinetic theory, based upon the idea that thermal
equilibrium is achieved by the energy exchange of point masses upon
collisions. \ A single harmonic oscillator can be regarded as in thermal
equilibrium with a heat bath when we imagine the oscillator exchanging energy
with point particles. \ The point particles are free particles which exchange
energy upon collision with the other particles of the bath, and then exchange
energy upon collision with the mass of the dipole oscillator. \ In
equilibrium, the kinetic energy of the oscillator particle will match (on
average) the kinetic energy of the free particles providing the heat bath.
Thus the average energy of the oscillator in the heat bath provided by
particle collisions is directly related to the average kinetic energy of the
heat-bath particles and has nothing to do with the natural frequency
$\omega_{0}$ of the oscillator. \ The average kinetic energy (in one spatial
dimension) of the heat-bath particles is denoted by $(1/2)k_{B}T.$ \ Thus the
average \textit{kinetic} energy of the (one-dimensional) oscillator which is
in thermal equilibrium with the heat bath is also $(1/2)k_{B}T.$ \ Now
according to classical mechanical theory, the kinetic energy of a harmonic
oscillator is shared equally with the potential energy of the oscillator
motion (as noted above in Eq. (\ref{Eav})); in a heat bath the mechanical
motion is merely interrupted and is changed by the collisions with the
heat-bath particles. \ Since the average kinetic energy is equal to
$(1/2)k_{B}T,$ the average potential energy must also be $(1/2)k_{B}T$. \ Thus
the average total energy of the oscillator must be $(1/2)k_{B}T+(1/2)k_{B}%
T=k_{B}T.$ \ This is the basic physics underlying the usual energy
equipartition theorem of classical statistical mechanics.\cite{Morse}

\subsubsection{Probability Distribution in Classical Statistical Mechanics}

According to classical statistical mechanics, the probability distribution on
phase space for the oscillator displacement $x$ and momentum $p$ is given by
$P(x,p)dxdp$ where \cite{Morse2}
\begin{equation}
P(x,p)=\left(  \frac{\omega_{0}}{2\pi\mathcal{E}_{s}}\right)  \exp\left(
-\frac{p^{2}/(2m)+m\omega_{0}^{2}x^{2}/2}{\mathcal{E}_{s}}\right)
\label{prob1}%
\end{equation}
and where $\mathcal{E}_{s}=k_{B}T$ is the average energy of a single dipole
oscillator in classical statistical mechanics. \ The subscript $s$ on
$\mathcal{E}_{s}$ refers to \textquotedblleft statistical
mechanics.\textquotedblright\ \ The probability distribution in Eq.
(\ref{prob1}) can be written as a product of two distributions%
\begin{equation}
P(x,p)=P_{x}(x,\mathcal{E}_{s})P_{p}(p,\mathcal{E}_{s}) \label{probprod}%
\end{equation}
with
\begin{equation}
P_{x}(x,\mathcal{E}_{s})=\left(  \frac{m\omega_{0}^{2}}{2\pi\mathcal{E}_{s}%
}\right)  ^{1/2}\exp\left(  -\frac{m\omega_{0}^{2}x^{2}/2}{\mathcal{E}_{s}%
}\right)  \text{ \ } \label{probx}%
\end{equation}
and%
\begin{equation}
\text{\ }P_{p}(p,\mathcal{E}_{s})=\left(  \frac{1}{2\pi m\mathcal{E}_{s}%
}\right)  ^{1/2}\exp\left(  -\frac{p^{2}/(2m)}{\mathcal{E}_{s}}\right)  .
\label{probp}%
\end{equation}
The random variables $x$ and $p$ have independent probability distributions.
\ The average value of powers of $x$ and $p$ is given by
\begin{equation}
\left\langle x^{2k}p^{2l}\right\rangle =\frac{(2k)!(2l)!}{k!l!2^{k+l}}\left(
\frac{\mathcal{E}_{s}}{m\omega_{0}^{2}}\right)  ^{k}\left[  m\mathcal{E}%
_{s}\right]  ^{l} \label{avxp}%
\end{equation}
for all even powers, with vanishing average value for any odd powers of $x$ or
of $p$.

\subsection{Two Interacting Dipole Oscillators in a Heat Bath}

When there are two interacting dipole oscillators described by the Hamiltonian
in Eq. (\ref{Hint1}), then there are correlations between the oscillator
motions, and also there are van der Waals forces tending to pull the two
oscillators together. \ 

\subsubsection{Correlation Function $\left\langle x_{A}x_{B}\right\rangle $}

When the two electric dipole oscillators interact electrostatically through
their dipole moments, they become correlated just as calculated in Eq.
(\ref{xAxBcl}) above in the section on classical mechanics of coupled
oscillators. \ Here we have the same equipartition energy for the oscillators
$\mathcal{E}_{+}=\mathcal{E}_{-}=\mathcal{E}_{s}=k_{B}T,$ but the frequencies
of oscillation are different, so that corresponding to Eq. (\ref{xAxBcl})
\begin{align}
\left\langle x_{A}x_{B}\right\rangle  &  =\frac{\mathcal{E}_{+}}{2m\omega
_{+}^{2}}-\frac{\mathcal{E}_{-}}{2m\omega_{-}^{2}}\nonumber\\
&  =\frac{\mathcal{E}_{s}}{2m}\left(  \frac{1}{\omega_{0}^{2}-2e^{2}/(mR^{3}%
)}-\frac{1}{\omega_{0}^{2}+2e^{2}/(mR^{3})}\right) \nonumber\\
&  \approx\frac{\mathcal{E}_{s}}{2m\omega_{0}^{2}}\frac{4e^{2}}{m\omega
_{0}^{2}R^{3}}=\frac{2\alpha^{2}\mathcal{E}_{s}}{e^{2}R^{3}}=\frac{2\alpha
^{2}k_{B}T}{e^{2}R^{3}} \label{corrstat}%
\end{align}
where we have expanded in the approximation $2e^{2}/R^{3}<<m\omega_{0}^{2},$
and the last line introduces the static electric
polarizability\cite{Griffiths2} of the oscillators where
\begin{equation}
\alpha=e^{2}/(m\omega_{0}^{2}). \label{alpha}%
\end{equation}

\subsubsection{van der Waals Attraction Between the Oscillators Calculated
from Electrostatic Forces}

In electromagnetism classes, students are asked to calculate the electrostatic
attraction between dipole oscillators. \ Therefore as our first calculation of
unretarded van der van der Waals forces, we will use the electromagnetic
context with only the final correlation given by classical statistical
mechanics. \ The unretarded van der Waals force $\mathbf{F}_{B}$ on oscillator
$B$ due to oscillator $A$ corresponds to the electrostatic force of one dipole
upon the other\cite{Griffiths3} $\mathbf{F}_{\text{on}B}=(\widehat{i}%
ex_{B}\cdot\nabla)\mathbf{E}_{A}\mathbf{(r}_{B})$ where the electric field
$\mathbf{E}_{A}$ corresponds to the electric dipole field of oscillator $A$.
\ In the present case, all the forces are in the $x$-direction, corresponding
to the direction of orientation of the oscillators and to the direction of the
separation between the oscillators. \ Thus the average force on the electric
dipole $\widehat{i}ex_{B}$ at $\mathbf{r}_{B}$ due to the electric dipole
$\widehat{i}ex_{A}$ at $\mathbf{r}_{A}$ is given by the time average of the
electrostatic force $\left\langle \mathbf{F}_{\text{on}B}\right\rangle
=\widehat{i}F_{\text{on}B}$
\begin{equation}
F_{s\text{on}B}=\left\langle ex_{B}\frac{\partial}{\partial R}\left(
\frac{2ex_{A}}{R^{3}}\right)  \right\rangle =-6\frac{e^{2}}{R^{4}}\left\langle
x_{A}x_{B}\right\rangle . \label{FonB}%
\end{equation}
The needed correlation function was obtained in Eq. (\ref{corrstat}) giving%
\begin{align}
F_{s\text{on}B}  &  =-6\frac{e^{2}}{R^{4}}\left\langle x_{A}x_{B}\right\rangle
=-6\frac{e^{2}}{R^{4}}\frac{2e^{2}\mathcal{E}_{s}}{\left(  m\omega_{0}%
^{2}\right)  ^{2}R^{3}}\nonumber\\
&  =-12\alpha^{2}\frac{\mathcal{E}_{s}}{R^{7}}=-12\alpha^{2}\frac{k_{B}%
T}{R^{7}} \label{FonB2}%
\end{align}
where $\alpha=e^{2}/(m\omega_{0}^{2})$ corresponds to the electric
polarizability of the oscillator and where $\mathcal{E}_{s}=k_{B}T$ is the
average energy of an isolated single oscillator. \ The force can be regarded
as arising from a potential function $\mathcal{U}_{s}(R,T)$ as $F_{s\text{on}%
B}=-\partial\mathcal{U}_{s}(R,T)/\partial R,$ where $\mathcal{U}_{s}(R,T)$ is
given by
\begin{equation}
\mathcal{U}_{s}(R,T)\approx-2\alpha^{2}\frac{\mathcal{E}_{s}}{R^{6}}%
=-2\alpha^{2}\frac{k_{B}T}{R^{6}}. \label{potStat}%
\end{equation}

\subsubsection{van der Waals Force from Helmholtz Free Energy}

Although the electromagnetic point of view above is direct, one obtains
further insight from the fully statistical mechanic point of view which
teaches students to work with the Hamiltonian function. \ The potential
function $\mathcal{U}_{s}(R,T)$ in equation (\ref{potStat}) which provides the
van der Waals force between the oscillators does not correspond to the
internal energy $U$ of the interacting dipole oscillator system. \ The thermal
energy of the interacting dipole oscillator system is just $2\mathcal{E}%
_{s}=2k_{B}T$ which is the sum of the thermal energies of the two individual
oscillators and makes no reference to the distance $R$ separating the two
oscillators. \ On the other hand, the van der Waals force in Eq. (\ref{FonB2})
clearly depends strongly on the separation between the oscillators. \ We
recall that in nonrelativistic statistical mechanics, the kinetic energy of
the oscillator masses is determined by the collisions of the point masses
providing the thermal bath, and there is no dependence of this energy upon the
oscillator frequencies. \ When the separation between the oscillators changes
at constant temperature, the thermal internal energy $U$ of the system does
\textit{not} change, but rather the entropy $S$ of the situation changes,
corresponding to $dQ=TdS=dU+dW$ where $dQ$ is the heat added, and $dW$ is the
work done by the system. \ According to thermodynamics, the force between the
oscillators depends upon the change in Helmholtz free energy $\mathcal{F}.$

The partition function for the two interacting oscillators in Eq.
(\ref{Hint1}) is given by\cite{Morse4}%
\begin{equation}
Z_{s}=%
{\textstyle\int_{-\infty}^{\infty}}
dp_{A}%
{\textstyle\int_{-\infty}^{\infty}}
dp_{B}%
{\textstyle\int_{-\infty}^{\infty}}
dx_{A}%
{\textstyle\int_{-\infty}^{\infty}}
dx_{B}\exp\left(  -H/\mathcal{E}_{s}\right)  .
\end{equation}
If we introduce the normal coordinates $x_{\pm}$ and $p_{\pm}$ with
frequencies $\omega_{\pm}^{2}=\omega_{0}^{2}\mp2e^{2}/(mR^{3})$, then the
partition function becomes%
\begin{align}
Z_{s}(\omega_{0},R,T)  &  =%
{\textstyle\int_{-\infty}^{\infty}}
dp_{+}%
{\textstyle\int_{-\infty}^{\infty}}
dp_{-}%
{\textstyle\int_{-\infty}^{\infty}}
dx_{+}%
{\textstyle\int_{-\infty}^{\infty}}
dx_{-}\exp\left(  -H/\mathcal{E}_{s}\right) \nonumber\\
&  =(2\pi mk_{B}T)\left(  \frac{2\pi k_{B}T}{m\omega_{+}^{2}}\right)
^{1/2}\left(  \frac{2\pi k_{B}T}{m\omega_{-}^{2}}\right)  ^{1/2}=\frac{(2\pi
k_{B}T)^{2}}{\omega_{+}\omega_{-}}. \label{Zpart}%
\end{align}
Next we evaluate the Helmholtz free energy as%
\begin{equation}
\mathcal{F}_{s}(\omega_{0},R,T)\mathcal{=-}k_{B}T\ln Z_{s}=k_{B}T\left(
\frac{1}{2}\ln\omega_{+}^{2}+\frac{1}{2}\ln\omega_{-}^{2}\right)  -k_{B}%
T\ln[(2\pi k_{B}T)^{2}].
\end{equation}
Now we omit the terms which are independent of the oscillator separation $R,$
and we expand $\omega_{+}^{2}$ and $\omega_{-}^{2}$ appearing in Eq.
(\ref{omegpm}) about $\omega_{0}^{2}$ using $\ln(1+x)=x-x^{2}/2+x^{3}/3-....$
\ Then we have
\begin{align}
\ln\omega_{\pm}^{2}  &  =\ln\left(  \omega_{0}^{2}\mp\frac{2e^{2}}{mR^{3}%
}\right)  =\ln\omega_{0}^{2}+\ln\left(  1+\frac{\mp2e^{2}}{m\omega_{0}%
^{2}R^{3}}\right) \nonumber\\
&  =\ln\omega_{0}^{2}+\left(  \frac{\mp2e^{2}}{m\omega_{0}^{2}R^{3}}\right)
-\frac{1}{2}\left(  \frac{\mp2e^{2}}{m\omega_{0}^{2}R^{3}}\right)  ^{2}+....
\end{align}
\ The interparticle potential $\mathcal{U}_{s}(R,T)=$ $\mathcal{F}_{s}%
(\omega_{0},R,T)-const$ for the force between the two oscillators from the
Helmholtz free energy $\mathcal{F}_{s}(\omega_{0},R,T)$ then becomes
\begin{equation}
\mathcal{U}_{s}(R,T)\mathcal{=-}\frac{2\alpha^{2}k_{B}T}{R^{6}}=-\frac
{2\alpha^{2}\mathcal{E}_{s}}{R^{6}}, \label{Ur3}%
\end{equation}
where $\alpha=e^{2}/(m\omega_{0}^{2})$ and $\mathcal{E}_{s}=k_{B}T.$ \ This
result is exactly the same as that found from the electrostatic force
calculation in Eq. (\ref{potStat}). \ Note that $\mathcal{E}_{s}=k_{B}T$ here
is the average energy for one oscillator.

\section{Interacting Dipole Oscillators in Classical Electrodynamics}

\subsection{Dipole Oscillators in Classical Electrodynamics}

Electric dipole oscillators are a staple of classical electrodynamics in
connection with dispersion, radiation emission, and radiation scattering.
\ The \textit{mechanical} oscillator behavior is given by the same
Hamiltonians as appear in Eqs. (\ref{Ham1}) and (\ref{Hint1}). \ However, now
the dipole oscillator systems include not only electrostatic dipole-dipole
interactions, but also connections to radiation emission and absorption. \ 

\subsection{Contrasts Between Classical Mechanics and Classical
Electrodynamics}

\subsubsection{Mismatch Between Classical Mechanics and Classical
Electrodynamics}

Classical electrodynamics was developed in the 19th century in the same time
period as the development of classical statistical mechanics. \ However, it
was noted that the two theories had contrasting and conflicting aspects.
\ Electrodynamics incorporated a fundamental velocity $c$ (derivable from the
constants of electrostatics and magnetostatics), whereas classical mechanics
(which is the basis for classical statistical mechanics) had no such
fundamental velocity. \ In complete contrast with classical mechanics, there
is no such thing as energy transfer due to a point collision within classical
electrodynamics. \ Sudden collisions of charged particles would involve large
energy losses to radiation. \ Indeed, classical electrodynamics involves
long-range Coulomb forces which do not fit into classical statistical
mechanics. \ Energy transfer in electromagnetic systems involves forces from
electromagnetic fields associated with charged particles or electromagnetic
waves. \ In complete contrast to the situation in classical statistical
mechanics, electric dipole oscillators of different frequencies interacting
through electromagnetic fields are \textit{not} forced to have the same
average energy in situations of steady-state behavior.

\subsubsection{Classical Electromagnetic Radiation Equilibrium}

At the end of the 19th century and beginning of the 20th century, the mismatch
between classical mechanics and classical electrodynamics led to attempts to
pin down the inertial frame of the \textquotedblleft luminiferous
ether\textquotedblright\ and also to attempts to understand the spectrum of
classical radiation equilibrium (the blackbody problem). \ By the early years
of the 20th century, it was realized that classical electrodynamics was a
relativistic theory satisfying Lorentz transformations whereas nonrelativistic
mechanics satisfied Galilean transformations. \ Furthermore, nonrelativistic
classical statistical mechanics did not fit with the blackbody radiation
spectrum except at low frequencies. \ \ 

\ It was Planck who recognized that, in equilibrium, the average energy of a
charged harmonic oscillator matched that of the random radiation at the
oscillator frequency.\cite{Lavenda} \ This possible dependence of the
oscillator kinetic energy on the frequency of the oscillator motion is totally
different from the kinetic-theory idea that the equilibrium kinetic energy
must be the same for all systems. \ However, around 1900, the importance of
relativity was not recognized and the possibility of classical zero-point
radiation was not considered. \ Indeed, the influence of the idea of kinetic
energy equipartition from kinetic theory was so strong that there was little
thought that Nature might contain random classical electromagnetic radiation
which did not vanish at the absolute zero of temperature. The physicists in
the early 20th century applied the ideas of nonrelativistic classical
statistical mechanics to the modes of electromagnetic radiation and claimed
that classical physics led inevitably to the Rayleigh-Jeans spectrum for
classical electromagnetic radiation in thermal equilibrium. \ Today we are
aware that this claim is wrong,\cite{hist} despite the fact that it is still
repeated in the textbooks of modern physics.\cite{Eisberg2}\ Like classical
statistical mechanics, quantum mechanics which developed during the early part
of the 20th century is a \textit{particle} theory which takes no account of
random radiation at zero temperature. \ The quantum theory of thermal
radiation is a \textit{particle} theory involving \textquotedblleft
photons\textquotedblright\ leading to a Planck spectrum which does not include
any zero-point radiation.\cite{Eisberg3}

Today we are much more aware of the importance of special relativity and of
the presence of random classical radiation even at zero temperature.
\ Classical zero-point radiation (which is a recognized possibility in
classical electrodynamics) is required in order to account for Casimir forces
within classical electromagnetism.\cite{B2011}\cite{under} \ Today we know
that relativistic classical electrodynamics which includes classical
electromagnetic zero-point radiation leads to classical thermal equilibrium at
the Planck spectrum including zero-point radiation,\cite{hist}\cite{rel}%
\cite{B2016} corresponding to an energy per normal mode given by%

\begin{equation}
\mathcal{E}_{Pzp}(\omega,T)=\frac{1}{2}\hbar\omega\coth\left(  \frac
{\hbar\omega}{2k_{B}T}\right)  =\frac{1}{2}\hbar\omega+\frac{\hbar\omega}%
{\exp[\hbar\omega/k_{B}T]-1}. \label{Planck1}%
\end{equation}
It is this spectrum of random radiation which we will use when discussing
thermal equilibrium for classical dipole oscillators within classical electrodynamics.

\subsection{Single Dipole Oscillator in Thermal Radiation}

\subsubsection{Radiation Energy Balance}

At thermal equilibrium, a dipole oscillator in classical electrodynamics must
be in equilibrium with the random radiation which surrounds it. \ Thus a
dipole oscillator must radiate away its energy, as calculated in classes \ in
electromagnetism, but it must also absorb energy from the random radiation
field which surrounds it. \ Indeed, it was Planck who first performed this
calculation involving thermal equilibrium for a dipole oscillator in random
classical radiation. \ Although Planck's calculation was important in the
history of physics, the behavior of a charged harmonic oscillator in random
classical radiation is unlikely to be familiar to advanced undergraduate
physics students today. \ \ Today our courses often do not introduce quantum
theory from a historical perspective. \ 

\subsubsection{Planck's Calculation}

Here we sketch the basic aspects of Planck's calculation.\cite{Lavenda}
\ Random classical radiation can be treated as source-free plane waves taken
as having periodic boundary conditions in a large cubic box of side $a$,%
\begin{equation}
\mathbf{E}_{\mathcal{R}}(\mathbf{r},t)=\operatorname{Re}%
{\textstyle\sum_{\mathbf{k}}}
{\textstyle\sum_{\lambda=1}^{2}}
\widehat{\epsilon}(\mathbf{k},\lambda)\left(  \frac{8\pi\mathcal{E}%
_{\mathcal{R}}(\omega)}{a^{3}}\right)  ^{1/2}\exp\{i[\mathbf{k}\cdot
\mathbf{r}-\omega t+\theta(\mathbf{k},\lambda)]\}, \label{Eran}%
\end{equation}%
\begin{equation}
\mathbf{B}_{\mathcal{R}}(\mathbf{r},t)=\operatorname{Re}%
{\textstyle\sum_{\mathbf{k}}}
{\textstyle\sum_{\lambda=1}^{2}}
\widehat{k}\times\widehat{\epsilon}(\mathbf{k},\lambda)\left(  \frac
{8\pi\mathcal{E}_{\mathcal{R}}(\omega)}{a^{3}}\right)  ^{1/2}\exp
\{i[\mathbf{k}\cdot\mathbf{r}-\omega t+\theta(\mathbf{k},\lambda)]\},
\label{Bran}%
\end{equation}
where the wave vector $\mathbf{k}$ takes the values $\mathbf{k}=\widehat{x}%
l2\pi/a+\widehat{y}m2\pi/a+\widehat{z}n2\pi/a$ for $l,m,n$ running over all
positive and negative integers, the constant $a$ refers to the length of a
side of the box for periodic boundary conditions with volume $V,$ $a^{3}=V,$
and the random phases $\theta(\mathbf{k},\lambda)$ are distributed uniformly
and independently over the interval $(0,2\pi].$\cite{Rice} \ The function
$\mathcal{E}_{\mathcal{R}}(\omega)$ is the electromagnetic energy in the
radiation normal mode of frequency $\omega=ck,$ and the subscript
$\mathcal{R}$ refers to \textquotedblleft random classical
radiation.\textquotedblright\ \ A small charged harmonic oscillator of mass
$m,$ natural (angular) frequency $\omega_{0},$ and charge $e$ located at
position $\mathbf{r}$ in this radiation has a classical equation of motion
following Newton's second law\cite{Griffiths4}
\begin{equation}
m\ddot{x}=-m\omega_{0}^{2}x+m\tau\dddot{x}+e\mathbf{E}_{rx}(\mathbf{r},t).
\label{xeq}%
\end{equation}
Here the forces involve the spring restoring force $-m\omega_{0}^{2}x$ of the
oscillator appearing in the Hamiltonian in Eq. (\ref{Ham1}), a radiation
damping force $m\tau\dddot{x}$ where
\begin{equation}
\tau=2e^{2}/(3mc^{3}),
\end{equation}
and the driving force $e\mathbf{E}_{r}(\mathbf{r},t)$ of the random electric
field in Eq. (\ref{Eran}), where $\mathbf{r}$ gives the location of the
(small) oscillator. \ Since we are dealing with linear systems, we can
consider separately the effects of \textit{one} of the (complex) plane waves
$\mathbf{E}_{\mathbf{k}\lambda}(\mathbf{r},t)$ in the sum appearing in Eq.
(\ref{Eran}),
\begin{equation}
\mathbf{E}_{\mathbf{k}\lambda}(\mathbf{r},t)=\widehat{\epsilon}(\mathbf{k}%
,\lambda)\left(  \frac{8\pi\mathcal{E}_{\mathcal{R}}(\omega)}{a^{3}}\right)
^{1/2}\exp\{i[\mathbf{k}\cdot\mathbf{r}-\omega t+\theta(\mathbf{k}%
,\lambda)]\}. \label{Ekl}%
\end{equation}
\ With this (complex) plane wave (\ref{Ekl}) as the source electric field, the
oscillator equation (\ref{xeq}) has the (complex) steady-state solution%
\begin{equation}
x(t)=\frac{e\epsilon_{x}(\mathbf{k},\lambda)}{m}\left(  \frac{8\pi
\mathcal{E}_{\mathcal{R}}(\omega)}{a^{3}}\right)  ^{1/2}\frac{\exp
[i\mathbf{k\cdot r}-i\omega t+i\theta(\mathbf{k},\lambda)]}{-\omega^{2}%
+\omega_{0}^{2}-i\tau\omega^{3}} \label{xkl}%
\end{equation}
Summing over all the plane waves of the random radiation in Eq. (\ref{Eran})
and taking the real part, we have the (real) displacement of the dipole
oscillator in random radiation given by
\begin{equation}
x(t)=%
{\textstyle\sum_{\mathbf{k}}}
{\textstyle\sum_{\lambda=1}^{2}}
\frac{e\epsilon_{x}(\mathbf{k},\lambda)}{m}\left(  \frac{2\pi\mathcal{E}%
_{\mathcal{R}}(\omega)}{a^{3}}\right)  ^{1/2}\left\{  \frac{\exp
[i\mathbf{k\cdot r}-i\omega t+i\theta(\mathbf{k},\lambda)]}{-\omega^{2}%
+\omega_{0}^{2}-i\tau\omega^{3}}+cc\right\}  \label{xoft}%
\end{equation}
where \textquotedblleft cc\textquotedblright\ stands for \textquotedblleft
complex conjugate.\textquotedblright\ \ \ The average values for
$x(t),~x^{2}(t),$ etc. are found by averaging over the random phases of the
radiation modes with%

\begin{equation}
\left\langle \exp[-i\omega t+i\theta(\mathbf{k},\lambda)]\exp[-i\omega
^{\prime}t+i\theta(\mathbf{k}^{\prime},\lambda^{\prime})]\right\rangle =0
\label{ran1}%
\end{equation}
and%
\begin{equation}
\left\langle \exp[-i\omega t+i\theta(\mathbf{k},\lambda)]\exp[i\omega^{\prime
}t-i\theta(\mathbf{k}^{\prime},\lambda^{\prime})]\right\rangle =\delta
_{\lambda,\lambda^{\prime}}\delta_{\mathbf{k,k}^{\prime}}. \label{ran2}%
\end{equation}
Thus for our example in Eq. (\ref{xoft}), we have the averages
\begin{equation}
\left\langle x(t)\right\rangle =0,\text{ \ \ \ \ }\left\langle \dot
{x}(t)\right\rangle =0,
\end{equation}%
\begin{equation}
\left\langle x^{2}(t)\right\rangle =%
{\textstyle\sum_{\mathbf{k}}}
{\textstyle\sum_{\lambda=1}^{2}}
\epsilon_{x}^{2}(\mathbf{k},\lambda)\left(  \frac{2\pi\mathcal{E}%
_{\mathcal{R}}(\omega)}{a^{3}}\right)  \frac{2e^{2}}{m^{2}\left[  \left(
-\omega^{2}+\omega_{0}^{2}\right)  ^{2}+\left(  \tau\omega^{3}\right)
^{2}\right]  }, \label{x2t}%
\end{equation}%
\begin{equation}
\left\langle \dot{x}^{2}(t)\right\rangle =%
{\textstyle\sum_{\mathbf{k}}}
{\textstyle\sum_{\lambda=1}^{2}}
\epsilon_{x}^{2}(\mathbf{k},\lambda)\left(  \frac{2\pi\mathcal{E}%
_{\mathcal{R}}(\omega)}{a^{3}}\right)  \frac{2e^{2}\omega^{2}}{m^{2}\left[
\left(  -\omega^{2}+\omega_{0}^{2}\right)  ^{2}+\left(  \tau\omega^{3}\right)
^{2}\right]  }, \label{xd2}%
\end{equation}
and the average oscillator energy$~$%
\begin{equation}
\mathcal{E}_{r}(\omega_{0})=m\left\langle \dot{x}^{2}(t)\right\rangle
/2+m\omega_{0}^{2}\left\langle x^{2}(t)\right\rangle /2, \label{Er}%
\end{equation}
becomes
\begin{equation}
\mathcal{E}_{r}(\omega_{0})=%
{\textstyle\sum_{\mathbf{k}}}
{\textstyle\sum_{\lambda=1}^{2}}
\epsilon_{x}^{2}(\mathbf{k},\lambda)\left(  \frac{2\pi\mathcal{E}%
_{\mathcal{R}}(\omega)}{a^{3}}\right)  \frac{e^{2}(\omega_{0}^{2}+\omega^{2}%
)}{m\left[  \left(  -\omega^{2}+\omega_{0}^{2}\right)  ^{2}+\left(  \tau
\omega^{3}\right)  ^{2}\right]  }. \label{E4}%
\end{equation}
If the box for the periodic boundary conditions is taken as large, then the
normal modes are closely spaced, and the sums over $\mathbf{k}$ can be
replaced by integrals in $d^{3}k,$ $%
{\textstyle\sum_{\mathbf{k}}}
\rightarrow(a/2\pi)^{3}%
{\textstyle\int}
d^{3}k$ so that%
\begin{equation}
\left\langle x^{2}(t)\right\rangle =(a/2\pi)^{3}%
{\textstyle\int}
d^{3}k%
{\textstyle\sum_{\lambda=1}^{2}}
\epsilon_{x}^{2}(\mathbf{k},\lambda)\left(  \frac{2\pi\mathcal{E}%
_{\mathcal{R}}(\omega)}{a^{3}}\right)  \frac{2e^{2}}{m^{2}\left[  \left(
-\omega^{2}+\omega_{0}^{2}\right)  ^{2}+\left(  \tau\omega^{3}\right)
^{2}\right]  },
\end{equation}
with analogous expressions for $\left\langle \dot{x}^{2}(t)\right\rangle $ and
$\mathcal{E}_{r}(\omega_{0}).$ \ Summing over polarizations $\lambda,$ we have%
\begin{equation}%
{\textstyle\sum_{\lambda=1}^{2}}
\epsilon_{x}^{2}(\mathbf{k},\lambda)=1-\frac{k_{x}^{2}}{k^{2}}. \label{pol}%
\end{equation}
Then the angular integrations in $\mathbf{k}$ give%
\begin{equation}%
{\textstyle\int_{0}^{2\pi}}
d\phi%
{\textstyle\int_{0}^{\pi}}
d\theta\sin\theta(1-\cos^{2}\theta)=\frac{8\pi}{3}. \label{ang1}%
\end{equation}
\ In order to evaluate the final integrals over the magnitude of $k$, we
assume that the charge $e$ is small so that the damping parameter $\tau
\omega_{0}^{2}<<\omega_{0}.$ \ Then the integrands in Eqs. (\ref{x2t}%
)-(\ref{E4}) are sharply peaked. \ We replace every appearance of the
frequency $\omega=ck$ by $\omega_{0}$ except for the combination
$\omega-\omega_{0},$ we take the limits of the integral as running from
$-\infty$ to $+\infty,$ and we use%
\begin{equation}
\int_{-\infty}^{\infty}\frac{du}{a^{2}u^{2}+b^{2}}=\frac{\pi}{ab}.
\label{peak}%
\end{equation}
Then we find%
\begin{equation}
\left\langle x^{2}(t)\right\rangle =\frac{\mathcal{E}_{\mathcal{R}}(\omega
_{0})}{m\omega_{0}^{2}},\text{ \ }\left\langle \dot{x}^{2}(t)\right\rangle
=\frac{\mathcal{E}_{\mathcal{R}}(\omega_{0})}{m},\text{ } \label{aval}%
\end{equation}
and from Eq. (\ref{Er}) we have
\[
\mathcal{E}_{r}(\omega_{0})=\mathcal{E}_{\mathcal{R}}(\omega_{0}).
\]
We find that the average energy $\mathcal{E}_{r}(\omega_{0})$ of the
oscillator of natural frequency $\omega_{0}$ in random radiation is the same
as the average energy $\mathcal{E}_{\mathcal{R}}(\omega_{0})$\ of the
radiation normal mode at the frequency $\omega_{0}.$ This is Planck's historic
calculation of the average energy of an oscillator immersed in random
classical radiation.\cite{Lavenda} \ In a similar fashion, we can start with
the expression (\ref{xoft}) and evaluate averages for general products of
positions and momenta. \ We find\cite{B1975b} that the average values of the
positions and momenta correspond to%
\begin{equation}
\left\langle x^{2k}p^{2l}\right\rangle =\frac{(2k)!(2l)!}{k!l!2^{k+l}}\left(
\frac{\mathcal{E}_{r}(\omega_{0})}{m\omega_{0}^{2}}\right)  ^{k}\left[
m\mathcal{E}_{r}(\omega_{0})\right]  ^{l} \label{avr}%
\end{equation}
with all average values involving odd powers vanishing. \ This is the same
form as Eq. (\ref{avxp}) found from nonrelativistic classical statistical
mechanics; however, in the electromagnetic case here, the average energy
$\mathcal{E}_{r}(\omega_{0})$ may depend on the frequency $\omega_{0}$ of the
oscillator whereas in the classical statistical mechanical case the average
energy $\mathcal{E}_{s}=k_{B}T$ is independent of $\omega_{0}.$\ 

\subsubsection{Surprising Aspects of the Radiation Balance}

The classical electrodynamics description may be regarded as surprising since
the charge $e$ which coupled the oscillator to the random classical radiation
does not appear in the final results Eqs. (\ref{aval}) and (\ref{avr}). \ It
turns out that \textit{any} electromagnetic coupling for the oscillator (for
example a quadrupole coupling involving two positive charges at the end of a
spring) will give the same zero-coupling limit.\cite{B1975a} \ Even in the
zero-coupling limit, the influence of the random radiation is still evident in
the non-vanishing random behavior of the classical oscillator. \ The
independence from the details of the electromagnetic interaction arises since
both the energy loss and the energy pick-up from the random radiation are
increased or decreased in the same fashion as the electromagnetic interaction
is altered.

It is clear from the results in Eqs. (\ref{avxp}), and (\ref{avr}), that both
classical descriptions of the single harmonic oscillator involve a Gaussian
probability distribution%
\begin{equation}
P(x,p)=\left(  \frac{\omega_{0}}{2\pi\mathcal{E}}\right)  \exp\left(
-\frac{p^{2}/(2m)+m\omega_{0}^{2}x^{2}/2}{\mathcal{E}}\right)  =P_{x}%
(x,\mathcal{E})P_{p}(p,\mathcal{E}) \label{probcl}%
\end{equation}
with different assumptions regarding the average oscillator energy
$\mathcal{E}.$ In kinetic theory and in classical statistical mechanics, the
average kinetic energy for each particle is the same in equilibrium;
equilibrium kinetic energy is completely independent from any oscillation
frequency of the particle. \ In contrast, electromagnetic systems in random
radiation do not show this kinetic-energy equipartition; rather, the particle
kinetic energy (of the oscillator) depends upon the random energy
$\mathcal{E}_{\mathcal{R}}(\omega_{0})$\ in the radiation at the natural
frequency $\omega_{0}$ of the oscillator, and can be quite different for
oscillators of different frequencies.

\subsubsection{Limitations of Dipole Oscillator Systems}

There are two aspects of a dipole oscillator which we wish to emphasize. \ The
first is that a point dipole oscillator can be regarded as the low-velocity
limit of a relativistic oscillator system. \ This allows us to consider point
dipole oscillators within relativistic classical electrodynamics. \ This
small-oscillator low-velocity limit has been discussed at length.\cite{B2016}
\ Second, a dipole oscillator does not determine the frequency spectrum of
thermal radiation. \ The oscillator energy will \textit{match} the energy of
the random radiation but does not \textit{determine} the spectrum. \ A dipole
oscillator does not act as a \textquotedblleft black
particle\textquotedblright\ in rescattering radiation into different
\textit{frequencies}. \ Rather a harmonic oscillator is a linear system which
will scatter radiation into different \textit{directions}, tending to make the
radiation pattern more nearly isotropic;\cite{B1975a} however, the scattered
radiation is of exactly the same frequency as the incident radiation. \ Thus a
point dipole oscillator does not push radiation toward the equilibrium
frequency spectrum of thermal radiation.\ \ It was precisely because of this
failure of a point oscillator to act as a black particle that Planck turned to
statistical mechanics applied to the oscillator in trying to determine the
radiation spectrum of thermal equilibrium.\cite{Lavenda}

\ However, if one applies \textit{nonrelativistic} classical statistical
mechanics to the oscillator (as in the modern physics texts\cite{Eisberg2}) or
uses a \textit{nonrelativistic} nonlinear classical scatterer (as has been
done in the research literature\cite{nonlin}), then one indeed finds the
Rayleigh-Jeans spectrum. \ Only by going to relativistic considerations
including Lorentz-invariant classical zero-point radiation, does one find the
Planck spectrum with zero-point radiation $\mathcal{E}_{Pzp}\mathcal{(}%
\omega,T)$ of Eq. (\ref{Planck1}) as the equilibrium spectrum of classical
physics.\cite{hist}\cite{rel}\cite{B2016} \ The Planck spectrum for thermal
radiation depends crucially upon relativity within classical physics.

\subsection{Two Interacting Dipole Oscillators}

\subsubsection{Oscillator Motion in Random Classical Radiation}

When two electric dipole harmonic oscillators are located in random classical
electromagnetic radiation $\mathbf{E}_{\mathcal{R}}(\mathbf{r},t)$, they are
driven into random oscillation.\ The oscillator equations of motion take the
form
\begin{align}
m\ddot{x}_{A}  &  =-m\omega_{0}^{2}x_{A}+m\tau\dddot{x}_{A}+eE_{Bx}%
(\mathbf{r}_{A},t)+eE_{\mathcal{R}x}(\mathbf{r}_{A},t)\nonumber\\
m\ddot{x}_{B}  &  =-m\omega_{0}^{2}x_{B}+m\tau\dddot{x}_{B}+eE_{Ax}%
(\mathbf{r}_{B},t)+eE_{\mathcal{R}x}(\mathbf{r}_{B},t) \label{xeqs}%
\end{align}
where $E_{Bx}(\mathbf{r}_{A},t)$ is the $x$-component of the electric dipole
field of oscillator $B$ at the location of oscillator $A$, and $E_{Ax}%
(\mathbf{r}_{B},t)$ is the analogous dipole field due to oscillator $A.$
\ Again for point dipole oscillators, this is a system of linear equations,
and accordingly we can treat separately the contribution (\ref{Ekl}) from each
plane wave $\mathbf{E}_{\mathbf{k}\lambda}$ appearing in the random radiation
sum of Eq. (\ref{Eran}). \ Introducing $\mathbf{E}_{\mathbf{k}\lambda
}(\mathbf{r},t)$ from Eq. (\ref{Ekl}) into the equations (\ref{xeqs}), the
(complex) steady-state equations become%
\begin{align}
-\omega^{2}mx_{A}  &  =-m\omega_{0}^{2}x_{A}+i\omega^{3}m\tau x_{A}%
-m\eta(k,R)x_{B}+eE_{\mathbf{k\lambda}x}(\mathbf{r}_{A},t)\nonumber\\
-\omega^{2}mx_{B}  &  =-m\omega_{0}^{2}x_{B}+i\omega^{3}m\tau x_{B}%
-m\eta(k,R)x_{A}+eE_{\mathbf{k\lambda}x}(\mathbf{r}_{B},t) \label{xeqs2}%
\end{align}
where
\begin{equation}
\eta(k,R)=-\frac{2e^{2}}{mR^{3}}\left(  1-ikR\right)  \exp[ikR]. \label{eta1}%
\end{equation}
The terms involving $\eta$ give the full electric fields of the oscillating
dipoles (and not just the electrostatic field) at the position of the other
dipole. \ 

If we divide through by $m,$ and then add and subtract the two equations in
(\ref{xeqs2}) while dividing by $\sqrt{2}$, we can deal with the normal modes
$x_{+}~$\ and $x_{-}$ introduced in Eqs. (\ref{norm}) so that we have%
\begin{align}
-\omega^{2}x_{+}  &  =-\omega_{0}^{2}x_{+}+i\omega^{3}\tau x_{+}%
-\eta(k,R)x_{+}+(e/m)\left[  E_{\mathbf{k\lambda}x}(\mathbf{r}_{A}%
,t)+E_{\mathbf{k\lambda}x}(\mathbf{r}_{B},t)\right]  /\sqrt{2}\nonumber\\
-\omega^{2}x_{-}  &  =-\omega_{0}^{2}x_{-}+i\omega^{3}\tau x_{-}%
+\eta(k,R)x_{-}+(e/m)\left[  E_{\mathbf{k\lambda}x}(\mathbf{r}_{A}%
,t)-E_{\mathbf{k\lambda}x}(\mathbf{r}_{B},t)\right]  /\sqrt{2} \label{xeqs3}%
\end{align}
and
\begin{equation}
x_{\pm}=\frac{(e/m)\left[  E_{\mathbf{k\lambda}x}(\mathbf{r}_{A},t)\pm
E_{\mathbf{k\lambda}x}(\mathbf{r}_{B},t)\right]  /\sqrt{2}}{-\omega^{2}%
+\omega_{0}^{2}-i\omega^{3}\tau\pm\eta(k,R)}. \label{xeqs4}%
\end{equation}
Reintroducing the full sum over all plane waves in the random radiation
(\ref{Eran}), we find the (real) normal modes%
\begin{align}
x_{\pm}(t)  &  =%
{\textstyle\sum_{\mathbf{k}}}
{\textstyle\sum_{\lambda=1}^{2}}
\frac{e\epsilon_{x}(\mathbf{k},\lambda)}{m}\left(  \frac{2\pi\mathcal{E}%
_{\mathcal{R}}(\omega)}{a^{3}}\right)  ^{1/2}\nonumber\\
&  \times\left\{  \frac{\{\exp[i\mathbf{k\cdot r}_{A}]\pm\exp[i\mathbf{k}%
\cdot\mathbf{r}_{B}]\}\exp[-i\omega t+i\theta(\mathbf{k},\lambda)]/\sqrt{2}%
}{-\omega^{2}+\omega_{0}^{2}-i\tau\omega^{3}\pm\eta(k,R)}+cc\right\}  .
\label{xpm}%
\end{align}
Now squaring, and averaging over the random phases $\theta(\mathbf{k}%
,\lambda)$ as in Eqs. (\ref{ran1}) and (\ref{ran2}), we find
\begin{align}
\left\langle x_{\pm}^{2}(t)\right\rangle  &  =%
{\textstyle\sum_{\mathbf{k}}}
{\textstyle\sum_{\lambda=1}^{2}}
\frac{e^{2}}{m^{2}}\epsilon_{x}^{2}(\mathbf{k},\lambda)\left(  \frac
{2\pi\mathcal{E}_{\mathcal{R}}(\omega)}{a^{3}}\right)  \left\{  \left\vert
\frac{\{\exp[i\mathbf{k}\cdot\mathbf{r}_{A}]\pm\exp[i\mathbf{k}\cdot
\mathbf{r}_{B}]\}}{\left\vert -\omega^{2}+\omega_{0}^{2}-i\tau\omega^{3}%
\pm\eta(k,R)\right\vert ^{2}}\right\vert ^{2}\right\} \nonumber\\
&  =%
{\textstyle\sum_{\mathbf{k}}}
{\textstyle\sum_{\lambda=1}^{2}}
\frac{e^{2}}{m^{2}}\epsilon_{x}^{2}(\mathbf{k},\lambda)\left(  \frac
{2\pi\mathcal{E}_{\mathcal{R}}(\omega)}{a^{3}}\right)  \left\{  2\frac
{\{1\pm\cos[\mathbf{k}\cdot(\mathbf{r}_{A}-\mathbf{r}_{B})]\}}{\left\vert
-\omega^{2}+\omega_{0}^{2}-i\tau\omega^{3}\pm\eta(k,R)\right\vert ^{2}%
}\right\}  . \label{xpm3}%
\end{align}
\ Again assuming that the box for the periodic boundary conditions is taken as
large so that the normal modes are closely spaced, then the sums over
$\mathbf{k}$ can be replaced by integrals in $d^{3}k,$ $%
{\textstyle\sum_{\mathbf{k}}}
\rightarrow(a/2\pi)^{3}%
{\textstyle\int}
d^{3}k$. \ The sum over polarization is the same as in Eq. (\ref{pol}). \ Next
we carry out the angular integrations in $\mathbf{k}$ as \
\begin{align}
&
{\textstyle\int_{0}^{2\pi}}
d\phi%
{\textstyle\int_{0}^{\pi}}
d\theta\sin\theta(1-\cos^{2}\theta)(1\pm\cos kR\cos\theta)\nonumber\\
&  =2\pi%
{\textstyle\int_{-1}^{1}}
dx(1-x^{2})(1\pm\cos kRx)\nonumber\\
&  =8\pi\left(  \frac{1}{3}\mp\frac{\cos(kR)}{(kR)^{2}}\pm\frac{\sin
(kR)}{(kR)^{3}}\right) \nonumber\\
&  =\frac{4\pi}{(e^{2}/m)k^{3}}\operatorname{Im}(i\tau\omega^{3}\mp\eta)
\end{align}
Then the expression (\ref{xpm3}) for the displacement squared collapses to%

\begin{equation}
\left\langle x_{\pm}^{2}(t)\right\rangle =%
{\textstyle\int_{k=0}^{k=\infty}}
dk\left(  \frac{2\mathcal{E}_{\mathcal{R}}(\omega)}{\pi mk}\right)  \left\{
\frac{\operatorname{Im}(i\tau\omega^{3}\mp\eta)}{[-\omega^{2}+\omega_{0}%
^{2}+\operatorname{Re}\eta]^{2}+[\operatorname{Im}(i\tau\omega^{3}\mp
\eta)]^{2}}\right\}
\end{equation}
We are assuming that the interaction between the oscillators is small so that
the frequency shift associated with $\operatorname{Re}\eta(\omega_{0}/c,R)$ is
small compared with the natural oscillator frequency $\omega_{0},$
\begin{equation}
\operatorname{Re}\eta(\omega_{0}/c,R)=-\frac{2e^{2}}{mR^{3}}[\cos
(kR)+kR\sin(kR)]<<\omega_{0}^{2}.
\end{equation}
Then the integrand is strongly peaked at the resonant angular frequency
$\omega_{\pm}$ with
\begin{align}
\omega_{\pm}^{2}  &  =\omega_{0}^{2}\pm\operatorname{Re}\eta(\omega
_{0}/c,R)\nonumber\\
\omega_{\pm}  &  \approx\omega_{0}\pm\frac{1}{2\omega_{0}}\operatorname{Re}%
\eta(\omega_{0}/c,R).
\end{align}
Inserting $\omega_{\pm}$ for every frequency which does not involve the
difference $\omega-\omega_{\pm},$ and extending the integrations from
$-\infty~$to $+\infty,$ we have%

\begin{equation}
\left\langle x_{\pm}^{2}(t)\right\rangle =%
{\textstyle\int_{\omega=-\infty}^{\omega=\infty}}
d\omega\left(  \frac{2\mathcal{E}_{\mathcal{R}}(\omega_{\pm})}{\pi
m\omega_{\pm}}\right)  \left\{  \frac{\operatorname{Im}[i\tau\omega_{\pm}%
^{3}\mp\eta(\omega_{\pm}/c,R)]}{4\omega_{\pm}^{2}(\omega-\omega_{\pm}%
)^{2}+\{\operatorname{Im}[i\tau\omega_{\pm}^{3}\mp\eta(\omega_{\pm
}/c,R)]\}^{2}}\right\}  \label{xpms}%
\end{equation}
Now using the definite integral in Eq. (\ref{peak}), our expression becomes%
\begin{equation}
\left\langle x_{\pm}^{2}(t)\right\rangle =\left(  \frac{2\mathcal{E}%
_{\mathcal{R}}(\omega_{\pm})}{\pi m\omega_{\pm}}\right)  \left\{  \frac{\pi
}{2\omega_{\pm}}\right\}  =\frac{\mathcal{E}_{\mathcal{R}}(\omega_{\pm}%
)}{m\omega_{\pm}^{2}}.
\end{equation}
Proceeding in a similar fashion and setting $\mathcal{E}_{r}(\omega_{\pm
})=m\left\langle \dot{x}_{\pm}^{2}\right\rangle /2+m\omega_{0}^{2}\left\langle
x_{\pm}^{2}\right\rangle /2$, we obtain
\begin{equation}
\left\langle x_{\pm}^{2}(t)\right\rangle =\frac{\mathcal{E}_{\mathcal{R}%
}(\omega_{\pm})}{m\omega_{\pm}^{2}},\text{ \ }\left\langle \dot{x}_{\pm}%
^{2}(t)\right\rangle =\frac{\mathcal{E}_{\mathcal{R}}(\omega_{\pm})}{m},\text{
and }\mathcal{E}_{r}(\omega_{\pm})=\mathcal{E}_{\mathcal{R}}(\omega_{\pm}).
\end{equation}
These expressions take the same form as in Eq. (\ref{aval}) found for a single
dipole oscillator in random classical electromagnetic radiation. \ In the
unretarded approximation, $kR<<1,$ the resonant frequencies become%
\begin{equation}
\omega_{\pm}^{2}=\omega_{0}^{2}\pm\operatorname{Re}\eta(\omega_{0}%
/c,R)\approx\omega_{0}^{2}\mp\frac{2e^{2}}{mR^{3}} \label{ws}%
\end{equation}
and \
\begin{equation}
\omega_{\pm}\approx\left(  \omega_{0}^{2}\mp\frac{2e^{2}}{mR^{3}}\right)
^{1/2}\approx\omega_{0}\left[  1\mp\frac{e^{2}}{m\omega_{0}^{2}R^{3}}-\frac
{1}{2}\left(  \frac{e^{2}}{m\omega_{0}^{2}R^{3}}\right)  ^{2}\right]
\label{ww}%
\end{equation}
with%
\begin{equation}
\omega_{+}-\omega_{-}\approx-\frac{2e^{2}}{m\omega_{0}R^{3}}. \label{odiff}%
\end{equation}

\subsubsection{Correlation Function $\left\langle x_{A}x_{B}\right\rangle $}

The correlation function $\left\langle x_{A}x_{B}\right\rangle $ takes a form
similar to that found above in Eq. (\ref{corrstat}), here becoming%

\begin{equation}
\left\langle x_{A}x_{B}\right\rangle =\left\langle (x_{+}+x_{-})(x_{+}%
-x_{-})/2\right\rangle =\left\langle x_{+}^{2}-x_{-}^{2}\right\rangle
/2=\frac{\mathcal{E}_{r}(\omega_{+})}{2m\omega_{+}^{2}}-\frac{\mathcal{E}%
_{r}(\omega_{-})}{2m\omega_{-}^{2}} \label{corr4}%
\end{equation}
where now (in contrast to the situation in Eq. (\ref{corrstat})) the energy of
the normal mode can depend upon the frequency of the normal mode. \ If we
introduce the Planck spectrum with zero-point radiation given in Eq.
(\ref{Planck1}) and note Eq. (\ref{odiff}), then we find%
\begin{align}
\left\langle x_{A}x_{B}\right\rangle  &  =\frac{\mathcal{E}_{r}(\omega_{+}%
)}{2m\omega_{+}^{2}}-\frac{\mathcal{E}_{r}(\omega_{-})}{2m\omega_{-}^{2}%
}\approx\frac{(\omega_{+}-\omega_{-})}{2m}\frac{\partial}{\partial\omega
}\left(  \frac{\mathcal{E}_{r}(\omega)}{\omega^{2}}\right)  _{\omega
=\omega_{0}}\nonumber\\
&  =-\frac{2e^{2}}{m\omega_{0}R^{3}}\frac{1}{2m}\frac{\partial}{\partial
\omega}\left(  \frac{\hbar\omega\coth[\hbar\omega/(2k_{B}T)]}{2\omega^{2}%
}\right)  _{\omega=\omega_{0}}\nonumber\\
&  =-\frac{e^{2}}{m^{2}\omega_{0}R^{3}}\frac{\partial}{\partial\omega}\left(
\frac{\hbar\coth[\hbar\omega/(2k_{B}T)]}{2\omega}\right)  _{\omega=\omega_{0}%
}.
\end{align}
This correlation function for the two oscillators immersed in random radiation
is quite different from the correlation found in Eq. (\ref{corrstat}) for two
oscillators in a classical statistical mechanical heat bath precisely because
energy equipartition does not hold for thermodynamic equilibrium within
classical electrodynamics..

\subsubsection{Van der Waals Forces}

The average force on dipole $B$ is given by $\mathbf{F}_{\text{on}%
B}=\left\langle [(e\widehat{i}x_{B})\cdot\nabla_{B}]\mathbf{E}(\mathbf{r}%
_{B},t)\right\rangle .$ \ For oscillators oriented along the $x$-axis and
separated along the $x$-axis, this gives a force along the $x$-axis
$\mathbf{F}=\widehat{i}ex_{x}(\partial E_{x}/\partial x).$ \ The electric
field $E_{x}$ at dipole $B$ is the sum of the electric field $E_{Ax}$ arising
from the oscillating electric dipole $A$ and the source-independent random
radiation $E_{\mathcal{R}x}$. \ Thus the average force $\mathbf{F}%
_{\text{on}B}=\widehat{i}F_{\text{on}B}$ on oscillator $B$ is given by
\begin{align}
\mathbf{F}_{\text{on}B}  &  =\left\langle [(e\widehat{i}x_{B})\cdot\nabla
_{B}][\mathbf{E}_{\mathcal{R}r}(\mathbf{r}_{B},t)+\mathbf{E}_{A}%
\mathbf{(r}_{B},t)]\right\rangle \nonumber\\
&  =\widehat{i}\left\langle ex_{B}\frac{\partial}{\partial R}E_{\mathcal{R}%
x}(\mathbf{r}_{B},t)\right\rangle +\widehat{i}\left\langle ex_{B}%
\frac{\partial}{\partial R}E_{Ax}(\mathbf{r}_{B},t)\right\rangle
\end{align}
In the unretarded limit considered here, the contribution from the first term
vanishes; the first term gives a non-zero contribution only for the retarded
forces.\cite{B1973} \ The calculation of unretarded van der Waals forces
proceeds exactly as discussed above where we obtain the average electrostatic
force of one dipole upon the other using Eq. (\ref{FonB}). \ Thus here we have%
\begin{align}
F_{\text{on}B}  &  =\left\langle ex_{B}\frac{\partial}{\partial R}\left(
\frac{2ex_{A}}{R^{3}}\right)  \right\rangle =-6\frac{e^{2}}{R^{4}}\left\langle
x_{A}x_{B}\right\rangle \nonumber\\
&  =6\frac{e^{2}}{R^{4}}\frac{e^{2}}{m^{2}\omega_{0}R^{3}}\frac{\partial
}{\partial\omega}\left(  \frac{\hbar\coth[\hbar\omega/(2k_{B}T)]}{2\omega
}\right)  _{\omega=\omega_{0}}. \label{FonB3}%
\end{align}
The potential function $\mathcal{U}_{r}(R)$ associated with the van der Waals
force is
\begin{equation}
\mathcal{U}_{r}(R,T)=\frac{e^{4}}{m^{2}\omega_{0}R^{6}}\frac{\partial
}{\partial\omega}\left(  \frac{\hbar\coth[\hbar\omega/(2k_{B}T)]}{2\omega
}\right)  _{\omega=\omega_{0}}. \label{cedpot}%
\end{equation}
There are two special cases which are worth considering separately. \ 

\paragraph{High-Temperature Limit}

The first special case is the high-temperature limit, $\hbar\omega
_{0}/2<<k_{B}T,$ of the Planck spectrum with zero-point radiation
(\ref{Planck1}) where the energy goes over fully to the Rayleigh-Jeans form
and so the energy per normal mode is independent of the frequency,
$\mathcal{E}(\omega_{+})=\mathcal{E}(\omega_{+})=k_{B}T.$ \ In this case, we
need the expansion for $\coth x$ for small $x,~\coth x=1/x+x/3-x^{3}/45+...,$
so that
\begin{equation}
F_{r\text{on}B}\approx6\frac{e^{2}}{R^{4}}\frac{e^{2}}{m^{2}\omega_{0}R^{3}%
}\frac{\partial}{\partial\omega}\left(  \frac{\hbar}{2\omega}\frac{2k_{B}%
T}{\hbar\omega}\right)  _{\omega=\omega_{0}}=-12\alpha^{2}\frac{k_{B}T}{R^{7}%
}\text{ for }k_{B}T>>\hbar\omega_{0},
\end{equation}
with $\alpha=e^{2}/(m\omega_{0}^{2})$ corresponding to the electric
polarizability of the oscillator. \ The associated potential function
$\mathcal{U}_{r}$ for the force is
\begin{equation}
\mathcal{U}_{r}(R,T)=-2\alpha^{2}k_{B}T/R^{6},
\end{equation}
just as in Eq. (\ref{potStat}) for the case of nonrelativistic classical
statistical mechanics. \ 

\paragraph{Low-Temperature Limit}

The second special case is that of low temperature, $k_{B}T<<\hbar\omega_{0},$
where the Planck spectrum including classical zero-point radiation
(\ref{Planck1}) goes over to the zero-point spectrum, $\mathcal{E}_{zp}%
(\omega_{+})=(1/2)\hbar\omega_{+},$ $\mathcal{E}_{zp}(\omega_{-}%
)=(1/2)\hbar\omega_{-}.$ \ In this case, we need the limit of $\coth x$ for
large $x$, $\coth x\rightarrow1.$ \ Then the force between the two oscillators
is
\begin{equation}
F_{r\text{on}B}\approx6\frac{e^{2}}{R^{4}}\frac{e^{2}}{m^{2}\omega_{0}R^{3}%
}\frac{\partial}{\partial\omega}\left(  \frac{\hbar}{2\omega}\right)
_{\omega=\omega_{0}}=-6\frac{\alpha^{2}}{R^{7}}\frac{\hbar\omega_{0}}{2}\text{
\ for}~\hbar\omega_{0}>>k_{B}T,
\end{equation}
where $\alpha=e^{2}/(m\omega_{0}^{2})$ corresponds to the electric
polarizability of the oscillator. \ The force can be regarded as arising from
a potential function $\mathcal{U}_{r}(R,0)$ as $F_{r\text{on}B}=-\partial
\mathcal{U}_{r}(R,0)/\partial R,$ where $\mathcal{U}_{r}(R,0)$ is given by
\begin{equation}
\mathcal{U}_{r}(R,0)\approx-\frac{\alpha^{2}}{R^{6}}\mathcal{E}_{zp}%
(\omega_{0})=-\frac{\alpha^{2}}{R^{6}}\frac{\hbar\omega_{0}}{2}. \label{U0}%
\end{equation}

\subsection{ \ Zero-Point Energy in Relativistic Classical Electrodynamics}

We have seen here that classical electrodynamics allows both a
high-temperature and low-temperature limit for van der Waals forces. \ This
situation is in sharp contrast with nonrelativistic classical statistical
mechanics which has only one form for the van der Waals forces at all
temperatures. \ The nonrelativistic classical statistical mechanical result in
Eq. (\ref{Ur3}) vanishes at zero temperature where $\mathcal{E}_{s}%
=k_{B}T\rightarrow0.$ \ However, the general classical electrodynamic spectrum
$\mathcal{E}_{Pzp}(\omega,T)$ in Eq. (\ref{Planck1}) contains
temperature-independent random classical zero-point radiation which persists
even at zero-temperature. \ It is the random classical zero-point radiation
$\mathcal{E}_{zp}(\omega)$ which accounts for the van der Waals forces between
the dipole oscillators in the low-temperature limit of Eq. (\ref{U0}). \ 

Classical electromagnetic zero-point radiation corresponds to an energy
$\mathcal{E}_{zp}(\omega)=\hbar\omega/2$ per normal mode. \ Up to an over-all
multiplicative constant, this is the \textit{unique} \textit{spectrum} of
random radiation which is invariant under \textit{Lorentz transformation}%
.\cite{relinv} \ The zero-point radiation spectrum takes the same form in
every inertial frame and so has no preferred frame of reference. \ The
spectrum also corresponds to a divergent energy density, as must hold true for
a Lorentz-invariant spectrum. \ The zero-point radiation spectrum is quite
different from the thermal radiation spectrum at non-zero temperature which
has a preferred reference frame, namely the reference frame in which the
enclosing container is at rest. \ Furthermore, the thermal radiation
$\mathcal{E}_{Pzp}(\omega,T)-\mathcal{E}_{zp}(\omega)$ above the zero-point
radiation $\mathcal{E}_{zp}(\omega)$ must have a finite energy density which
is involved in thermodynamic relations.\cite{thermo} \ 

A nonrelativistic classical mechanical theory such as classical statistical
mechanics cannot support the idea of a zero-point energy because \textit{all}
kinetic energy is shared in collisions between point masses. \ On the other
hand, a point dipole oscillator immersed in classical zero-point radiation
will acquire the average energy of random radiation at the oscillator's
natural frequency; and this average energy includes zero-point energy. \ A
point classical oscillator can support the idea of a zero-point energy if it
shares its energy by interactions through classical electromagnetic forces. \ 

We should note that the zero-point energy of an oscillator $\mathcal{E}%
_{r}\mathcal{(}\omega_{0},0\mathcal{)}=\hbar\omega_{0}/2$ is adiabatic
invariant under a change in the natural frequency $\omega_{0\text{ }}$of the
oscillator, just as the spectrum of classical electromagnetic zero-point
energy is invariant under an adiabatic compression.\cite{thermo}

\section{Interacting Dipole Oscillators in Quantum Electrodynamics}

There are strong connections between quantum electrodynamics and classical
electrodynamics with classical electromagnetic zero-point radiation. \ If one
considers the interaction of point electric dipole oscillators with the
quantum electromagnetic field, the equations of motion in the Heisenberg
picture take the same form as the classical electrodynamic equations. \ Since
the equations are linear, one finds the same average values, now taken as
vacuum expectation values, for $\left\langle x_{A}^{2}\right\rangle
,~\left\langle x_{B}^{2}\right\rangle ,~\left\langle x_{A}x_{B}\right\rangle
,$ and indeed for all the quadratic expressions. \ Since quadratic expressions
are involved, the van der Waals force predicted by quantum electrodynamics and
by classical electrodynamics with classical zero-point radiation are the
same.\cite{VV} \ Indeed, there is general agreement at all temperatures
between classical electrodynamics with zero-point radiation and quantum
electrodynamics for free fields and for harmonic oscillator systems
\textit{provided all products of quantum operators are completely
symmetrized}.\cite{B1975b}

\section{Interacting Dipole Oscillators in Quantum Mechanics}

\subsection{Connection of Quantum Electrodynamics and Quantum Mechanics}

Quantum mechanics is often assumed to be a suitable limit of quantum
electrodynamics.\cite{B1975b} \ Because of the agreement between classical
electrodynamics and quantum electrodynamics for linear systems, we expect that
the quantum mechanics of harmonic oscillator systems should correspond to the
small-charge limit $e\rightarrow0$ of charged harmonic oscillators in
classical electrodynamics, \textit{provided all quantum operator products are
completely symmetrized. \ }

In the small-$e$ limit, a classical dipole oscillator in random classical
radiation takes the same distribution function as the random radiation at the
natural frequency of the oscillator. \ Thus in Eqs. (\ref{avr}) and
(\ref{probcl}) for a single oscillator in random radiation, there is no
dependence upon the charge $e$ associated with the dipole moment. \ 

Classical electrodynamics with classical electromagnetic zero-point radiation
is sometimes termed \textquotedblleft stochastic
electrodynamics.\textquotedblright\ \ The small-$e$ limit for the dipole
oscillators coupled to random zero-point radiation is sometimes termed
\textquotedblleft stochastic mechanics\textquotedblright\ in analogy with the
small-$e$ limit of quantum electrodynamics becoming quantum mechanics. For
harmonic oscillator systems, the results of stochastic mechanics agree with
the results of quantum mechanics provided all the quantum operator products
are completely symmetrized.\cite{B1975b}

\subsection{Single Oscillator}

\subsubsection{Harmonic Oscillator in Quantum Mechanics}

\ The quantum theory of the harmonic oscillator\cite{GriffithsQ} is familiar
to every advanced undergraduate physics student. \ The quantum Hamiltonian is
the same as that for the classical mechanical oscillator given in Eq.
(\ref{Ham1}), but the position $x$ and momentum $p$ now become quantum
operators $\hat{x}$ and $\hat{p}$. \ The energy eigenstates $|n>$ of the
oscillator are labeled by the integer index $n=0,1,2,...$ and correspond to
energies $\mathcal{E}_{q}(n)\mathcal{=}(n+1/2)\hbar\omega_{0}$. \ Here the
subscript $q$ on $\mathcal{E}_{q}(n)$ denotes \textquotedblleft
quantum.\textquotedblright\ \ The ground state corresponds to $n=0.$ \ 

\subsubsection{Contrasts Between Classical and Quantum Descriptions}

For the quantum oscillator, the vacuum expectation values of position-squared
$\left\langle 0\left\vert \hat{x}^{2}\right\vert 0\right\rangle $ and
momentum-squared $\left\langle 0\left\vert \hat{p}^{2}\right\vert
0\right\rangle $ correspond exactly to Eq. (\ref{aval}), provided
$\mathcal{E}_{\mathcal{R}}$ is replaced by $\mathcal{E}_{q}.\ $Indeed the
expectation values for powers of $\hat{x}$ alone or of powers of $\hat{p}$
alone give
\begin{equation}
\left\langle 0\left\vert \hat{x}^{2k}\right\vert 0\right\rangle =\frac
{(2k)!}{k!2^{k}}\left(  \frac{\mathcal{E}_{q}}{m\omega_{0}^{2}}\right)
^{k}\text{ \ and }\left\langle 0\left\vert \hat{p}^{2l}\right\vert
0\right\rangle =\frac{(2l)!}{l!2^{l}}\left[  m\mathcal{E}_{q}\right]  ^{l},
\end{equation}
vanish for all odd powers, and agree with the results of classical
electrodynamics with classical zero-point radiation given in Eq. (\ref{avr}).
$\ $However, in contrast to classical theories leading to independent
probability distributions for $x$ and $p$ as in Eqs. (\ref{avxp}) and
(\ref{avr}), the energy of the quantum operator corresponds to an eigenvalue,
so that average values involving $\hat{x}$ and $\hat{p}$ \textit{cannot} be
regarded as involving independent random variables. The quantum variables
cannot be regarded as described by a probability distribution such as given in
Eq. (\ref{probcl}) for the classical random oscillators. \ The quantum
description of the harmonic oscillator involves a ground state energy which
takes a unique value $\mathcal{E}_{q}(0)\mathcal{=}(0+1/2)\hbar\omega_{0}$
with no dispersion. \ The energy uncertainty in a state $|n>$ involves
$(\Delta\mathcal{E}_{q}(n))^{2}=\left\langle n|\hat{H}^{2}|n\right\rangle
-\left\langle n|\hat{H}|n\right\rangle ^{2}$ where $\left\langle n|\hat{H}%
^{2}|n\right\rangle =$ $\left\langle n|[\hat{p}^{2}/(2m)+m\omega_{0}^{2}%
\hat{x}^{2}/2]^{2}|n\right\rangle $, and so involves a very specific and
not-completely-symmetrized operator order (such as $\hat{p}^{2}\hat{x}%
^{2}+\hat{x}^{2}\hat{p}^{2}$) for the operators $\hat{x}$ and $\hat{p}.$

Indeed, it has been shown\cite{B1975b} that if one completely symmetrizes the
order of all factors in quantum operator products, then for free fields and
harmonic oscillator systems, at any temperature, quantum electrodynamics and
classic electrodynamics with classical zero-point radiation give the same
average values. \ There is \textit{disagreement }with such unsymmetrized
quantum operator expressions as $\left\langle 0|\hat{x}^{2}\hat{p}%
^{2}|0\right\rangle $ or $\left\langle 0|\hat{x}^{2}\hat{p}^{2}+\hat{p}%
^{2}\hat{x}^{2}|0\right\rangle /2,$ whereas the classical theory
\textit{agrees} with only the completely symmetrized expression $\left\langle
0|\hat{x}^{2}\hat{p}^{2}+\hat{x}\hat{p}\hat{x}\hat{p}++\hat{p}\hat{x}^{2}%
\hat{p}+\hat{x}\hat{p}^{2}\hat{x}+\hat{p}\hat{x}\hat{p}\hat{x}+\hat{p}^{2}%
\hat{x}^{2}|0\right\rangle /6.$ \ For example, the van der Waals forces agree
between the classical and the quantum theories, since these expressions are
bilinear, but the theories disagree regarding the dispersion. \ Thus the
Hamiltonian $\hat{H}=\hat{p}^{2}/(2m)+m\omega_{0}^{2}\hat{x}^{2}/2$ involves
completely symmetrized quantum operator products, but the expression for the
Hamiltonian squared $\hat{H}^{2}$\ does not. \ The classical theory does not
agree with the energy-eigenstate aspect of the quantum theory.

\subsubsection{Quantum Oscillator at Non-Zero Temperature}

At non-zero temperature $T>0,$ the expectation value of a quantum operator
$\hat{O}$ is given as an incoherent sum over excited states, so that the
quantum operator $\hat{O}$ takes the average value%
\begin{equation}
\left\langle \left\vert \hat{O}\right\vert \right\rangle _{T}=%
{\textstyle\sum_{n=0}^{\infty}}
\left\langle n\left\vert \hat{O}\right\vert n\right\rangle \frac{1}{Z_{q}}%
\exp\left(  \frac{-\mathcal{E}_{q}(n)}{k_{B}T}\right)  \label{qop}%
\end{equation}
where $Z_{q}$ is the quantum partition function for the system. \ The
thermodynamics of a quantum oscillator can be described by the partition
function\cite{Morse5}
\begin{equation}
Z_{q}(\omega_{0},T)=%
{\textstyle\sum_{n=0}^{\infty}}
\exp\left(  \frac{-\mathcal{E}_{q}(n)}{k_{B}T}\right)  =%
{\textstyle\sum_{n=0}^{\infty}}
\exp\left(  \frac{-\mathcal{(}n+1/2)\hbar\omega_{0}}{k_{B}T}\right)  =\frac
{1}{2}\text{csch}\left(  \frac{\hbar\omega_{0}}{2k_{B}T}\right)  .
\label{qpart}%
\end{equation}
The average quantum oscillator energy $\left\langle |H|\right\rangle _{T}$ at
non-zero temperature $T$ is given by
\begin{equation}
\mathcal{E}_{q}(T)=-\frac{\partial\ln Z_{q}(\omega_{0},T)}{\partial(1/k_{B}%
T)}=\frac{1}{2}\hbar\omega_{0}\coth\left(  \frac{\hbar\omega_{0}}{2k_{B}%
T}\right)  =\frac{1}{2}\hbar\omega_{0}+\frac{\hbar\omega_{0}}{\exp\left(
\hbar\omega_{0}/k_{B}T\right)  -1}, \label{EqT}%
\end{equation}
which is the same as the expression for a classical dipole oscillator in
thermal equilibrium in classical electrodynamics including zero-point
radiation as in Eq. (\ref{Planck1}). \ 

\subsubsection{High- and Low-Temperature Limits}

It is interesting that the quantum theory of the harmonic oscillator includes
the zero-point energy in Eq. (\ref{EqT}), whereas the photon description
giving the Planck spectrum $\mathcal{E}_{P}(\omega,T)$ does not include any
zero-point energy for the radiation.\cite{Eisberg3} \ In the limit of high
temperature for the Planck spectrum, we have
\begin{align}
\mathcal{E}_{P}(\omega,T)  &  =\frac{\hbar\omega_{0}}{\exp\left(  \hbar
\omega_{0}/k_{B}T\right)  -1}=\hbar\omega_{0}\left[  1+\frac{\hbar\omega_{0}%
}{k_{B}T}+\frac{1}{2}\left(  \frac{\hbar\omega_{0}}{k_{B}T}\right)
^{2}+...-1\right]  ^{-1}\nonumber\\
&  =k_{B}T-\frac{1}{2}\hbar\omega_{0}+O\left(  \frac{\hbar\omega_{0}}{k_{B}%
T}\right)  ,\text{ \ \ }k_{B}T>>\hbar\omega_{0}, \label{exp}%
\end{align}
which retains a finite correction $-\hbar\omega_{0}/2$ at high temperature
$T.$ \ Thus the average energy $\mathcal{E}_{q}(T)$ of the quantum oscillator
in Eq. (\ref{EqT}) goes over fully to the classical statistical mechanical
result $k_{B}T$ (without any nonvanishing correction at high $T)$ only if the
zero-point energy is included,%
\begin{align}
\mathcal{E}_{q}(T)  &  =\frac{1}{2}\hbar\omega_{0}+\frac{\hbar\omega_{0}}%
{\exp\left(  \hbar\omega_{0}/k_{B}T\right)  -1}\nonumber\\
&  =\frac{1}{2}\hbar\omega_{0}+\left[  k_{B}T-\frac{1}{2}\hbar\omega
_{0}+O\left(  \frac{\hbar\omega_{0}}{k_{B}T}\right)  \right] \nonumber\\
&  =k_{B}T+O\left(  \frac{\hbar\omega_{0}}{k_{B}T}\right)  ,\text{ \ \ }%
k_{B}T>>\hbar\omega_{0}. \label{EqTexp}%
\end{align}
Sometimes the need for the zero-point energy in order to achieve the classical
limit at high temperature is treated incorrectly in textbook
accounts.\cite{Morse6}

The difference between the average values $\left\langle \left\vert \hat
{O}\right\vert \right\rangle _{T}$ for unsymmetrized products of quantum
operators and the corresponding average values obtained in classical
electrodynamics with classical zero-point radiation persists at non-zero
temperatures; \ however, the fractional discrepancy becomes ever smaller as
the temperature becomes larger.\cite{B1975b} \ 

\subsection{Two Interacting Dipole Oscillators}

\subsubsection{van der Waals Forces at Zero Temperature}

The van der Waals forces between two dipole oscillators in quantum theory are
treated in a problem in a standard quantum mechanics
textbook.\cite{GriffithsQ} \ The unperturbed Hamiltonian corresponds to two
harmonic oscillators, and the interaction is that of two electric dipoles
$\widehat{i}ex_{A}$ and $\widehat{i}ex_{B}$ separated by a distance $R$ along
the $x$-axis, giving the interacting Hamiltonian in Eq. (\ref{Hint1}). When
the normal-mode coordinates $x_{+}$ and $x_{-}$ are introduced, the
Hamiltonian can be rewritten as in Eq. (\ref{Hint2}). \ Thus the Hamiltonian
can be viewed as that of two uncoupled harmonic oscillators of frequencies
$\omega_{\pm}=[\omega_{0}^{2}\mp2e^{2}/(mR^{3})]^{1/2}$ as in Eq.
(\ref{omegpm}), giving the energy eigenvalues $\mathcal{E}_{q}(k,l)$\ for the
two coupled oscillators as
\begin{equation}
\mathcal{E}_{q}(k,l)=(k+1/2)\hbar\omega_{+}+(l+1/2)\hbar\omega_{-}\text{ \ for
}k,l=0,1,2,....
\end{equation}
If the interaction between the oscillators is small $e^{2}/R^{3}<<m\omega
_{0}^{2}$, we may use the binomial expansion through second order in the
correction to the unperturbed ground state energy. \ The system ground state
energy is given by
\begin{align}
\mathcal{E}_{q}(0,0)  &  =\frac{1}{2}\hbar\left[  \left(  \omega_{0}^{2}%
+\frac{2e^{2}}{mR^{3}}\right)  ^{1/2}+\left(  \omega_{0}^{2}-\frac{2e^{2}%
}{mR^{3}}\right)  ^{1/2}\right] \nonumber\\
&  \approx\hbar\omega_{0}-\frac{1}{2}\left(  \frac{e^{2}}{m\omega_{0}^{2}%
}\right)  ^{2}\frac{\hbar\omega_{0}}{R^{6}}=2\mathcal{E}_{q}(\omega_{0}%
)-\frac{\alpha^{2}\mathcal{E}_{q}(\omega_{0})}{R^{6}}. \label{Uqmi}%
\end{align}
where $\mathcal{E}_{q}(\omega_{0})=\hbar\omega_{0}/2$ is the quantum ground
state energy for a single harmonic oscillator. \ 

The interaction energy $\mathcal{E}_{q}(0,0)-\hbar\omega_{0}$ (viewed as a
function of the inter-oscillator separation $R)$ provides a potential function
$\mathcal{U}_{q}(R,0)=-\alpha^{2}\mathcal{E}_{q}/R^{6}$ for the force between
the two dipole oscillators at zero temperature. \ We see that the quantum
mechanical ground-state energy in Eq. (\ref{Uqmi}) provides exactly the same
potential energy as the potential function in Eq. (\ref{U0}) found at
zero-temperature from classical electrodynamics with classical zero-point radiation.\ \ 

\subsubsection{van der Waals Forces at Non-Zero Temperature}

At non-zero temperature $T$, the Helmholtz free energy will provide the
potential function for the van der Waals forces between the oscillators. \ The
Helmholtz free energy $\mathcal{F}_{q}(\omega_{0},R,T)$ for the coupled
quantum system is analogous to that found for a single oscillator following
the partition function Eq. (\ref{qpart}), but now involving%
\begin{align}
Z_{q}(\omega_{0},R,T)  &  =%
{\textstyle\sum_{k,l=0}^{\infty}}
\exp\left(  \frac{-\mathcal{E}_{q}(k,l)}{k_{B}T}\right)  =%
{\textstyle\sum_{k,l=0}^{\infty}}
\exp\left(  -\frac{\mathcal{(}k+1/2)\hbar\omega_{+}+(l+1/2)\hbar\omega_{-}%
}{k_{B}T}\right) \nonumber\\
&  =\frac{1}{2}\text{csch}\left(  \frac{\hbar\omega_{+}}{2k_{B}T}\right)
\frac{1}{2}\text{csch}\left(  \frac{\hbar\omega_{-}}{2k_{B}T}\right)  ,
\label{part2}%
\end{align}
so that
\begin{align}
\mathcal{F}_{q}(\omega_{0},R,T)  &  =-k_{B}T\ln Z_{q}(\omega_{0}%
,R,T)\nonumber\\
&  =-k_{B}T\left\{  \ln\left[  \text{csch}\left(  \frac{\hbar\omega_{+}%
}{2k_{B}T}\right)  \right]  +\ln\left[  \text{csch}\left(  \frac{\hbar
\omega_{-}}{2k_{B}T}\right)  \right]  -\ln4\right\}  . \label{Fhel}%
\end{align}
We now carry out a Taylor series expansion for $f(\omega_{\pm})$ about the
argument $\omega_{0,}$ as $f(\omega_{\pm})=f(\omega_{0})+(\omega_{\pm}%
-\omega_{0})f^{\prime}(\omega_{0})+[(\omega_{\pm}-\omega_{0})^{2}%
/2]f^{\prime\prime}(\omega_{0})+....$, and note the value for $\omega_{\pm
}-\omega_{0}$ through order $1/R^{6}$ in Eq. (\ref{ww}). \ Then we find
\begin{align}
\mathcal{F}_{q}(\omega_{0},R,T)  &  \approx2\mathcal{F}_{q}(\omega
_{0},T)-k_{B}T\left(  \frac{-\hbar}{2k_{B}T}\right)  \coth\left(  \frac
{\hbar\omega_{0}}{2k_{B}T}\right)  \left[  \left(  \omega_{+}-\omega
_{0}\right)  +(\omega_{-}-\omega_{0})\right] \nonumber\\
&  -k_{B}T\left(  \frac{-\hbar}{2k_{B}T}\right)  \frac{\partial}%
{\partial\omega}\left\{  \coth\left(  \frac{\hbar\omega}{2k_{B}T}\right)
\right\}  _{\omega=\omega_{0}}\left[  \frac{\left(  \omega_{+}-\omega
_{0}\right)  ^{2}+(\omega_{-}-\omega_{0})^{2}}{2}\right] \nonumber\\
&  =2\mathcal{F}_{q}(\omega_{0},T)-k_{B}T\left(  \frac{-\hbar}{2k_{B}%
T}\right)  \coth\left(  \frac{\hbar\omega_{0}}{2k_{B}T}\right)  2\frac
{-\omega_{0}}{2}\left(  \frac{e^{2}}{m\omega_{0}^{2}R^{3}}\right)
^{2}\nonumber\\
&  -k_{B}T\left\{  \left(  \frac{-\hbar}{2k_{B}T}\right)  \frac{\partial
}{\partial\omega}\coth\left(  \frac{\hbar\omega}{2k_{B}T}\right)  \right\}
_{\omega=\omega_{0}}2\left(  \frac{e^{2}}{m\omega_{0}R^{3}}\right)  ^{2}%
\frac{1}{2} \label{Fq}%
\end{align}
The potential function $\mathcal{U}_{q}(R,T)$ for the van der Waals force
between the dipole oscillators can omit the terms in Eq. (\ref{Fq}) which are
independent of the spatial separation $R,$ and so takes the form
\begin{equation}
\mathcal{U}_{q}(R,T)=\frac{e^{4}}{m^{2}\omega_{0}R^{6}}\frac{\partial
}{\partial\omega}\left(  \frac{\hbar\coth[\hbar\omega/(2k_{B}T)]}{2\omega
}\right)  _{\omega=\omega_{0}}, \label{qedpot}%
\end{equation}
which is the same as that found in Eq. (\ref{cedpot}) from classical
electrodynamics including zero-point radiation. \ The classical electrodynamic
and quantum mechanical calculations for van der Waals forces agree at all temperatures.

\section{Closing Summary}

In this article, we have discussed harmonic oscillators within various
theoretical contexts of elementary physics. \ The classical mechanics of a
mass at the end of a spring teaches us that a linear oscillator shares its
energy equally between particle kinetic energy and spring potential energy.
\ Furthermore, oscillators weakly coupled by springs share their energy
between the oscillators, and can be described in terms of the normal modes of
oscillation of the oscillator system. \ Because atoms and molecules are often
described in terms of dipole oscillator systems, the physics of harmonic
oscillators reappears in discussions of van der Waals forces within the
contexts of classical statistical mechanics, classical electrodynamics,
quantum electrodynamics, and quantum mechanics. \ At high temperatures, the
van der Waals forces can be adequately described by nonrelativistic classical
statistical mechanics which arose from a picture of energy transfer between
the oscillators involving collisions between point masses which provide the
thermal bath for the oscillators. \ 

Classical electrodynamics has relativistic foundations which are very
different from those of nonrelativistic classical mechanics. \ However, point
dipole oscillators can be incorporated consistently into relativistic
classical electrodynamics. \ Classical electrodynamics gives correct
predictions\cite{B1975a} for Casimir forces, for van der Waals forces, for
oscillator specific heats, for blackbody radiation, and for diamagnetism only
if it incorporates Lorentz-invariant random classical zero-point radiation as
the source-free solution of Maxwell's equations. \ Because the
textbooks\cite{source} of classical electrodynamics do not include the
source-free radiation in the general solution to Maxwell's equations, many
teachers of physics do not realize that the possibility of source-free
radiation is an intrinsic part of classical electrodynamics, and that this
source-free radiation forms a boundary condition on Maxwell's differential
equations.\cite{contr} \ Classical electromagnetic theory must choose this
boundary condition so as to account for experimental observations. \ The
choice of this boundary condition is \textquotedblleft as much a part of the
postulates of the theory as the form of the Lagrangian or the value of the
electron charge.\textquotedblright\cite{Coleman} \ Relativistic classical
electrodynamics with classical zero-point radiation leads to Planck's spectrum
including zero-point radiation as the equilibrium spectrum of classical
thermal radiation. \ 

Any dipole oscillator in equilibrium with random radiation acquires the same
random energy as is present in the radiation normal modes of the same
frequency as the oscillator. \ Interacting dipole oscillators experience van
der Waals forces between the oscillators on account of the random radiation.
\ The van der Waals forces calculated from classical electrodynamics agree
with the results of classical statistical mechanics at high temperatures, and
agree with the results of quantum theory at all temperatures. \ 

Nonrelativistic classical statistical mechanics can be regarded as providing a
\textit{local} hidden variables theory for the behavior of interacting dipole
oscillators, with the colliding heat-bath particles providing the hidden
variables. \ This theory agrees with quantum theory only at high temperatures.
\ Classical electrodynamics with classical electromagnetic zero-point
radiation provides a classical theory which agrees with the results of quantum
theory for free fields and harmonic oscillator systems. \ The classical theory
can be regarded as providing a hidden variables theory for the behavior of
interacting dipole oscillators, with the random phases of the source-free
radiation modes providing the hidden variables. \ The classical electrodynamic
theory is\textit{ }not a \textit{local} hidden variables theory, but rather is
a \textit{nonlocal} theory involving correlations over finite distances. \ For
free fields and harmonic oscillator systems, the classical electrodynamic
forces agree with the quantum results at all temperatures. 

Acknowledgement

I wish to thank Professor Daniel C. Cole for his helpful comments on this manuscript.


\begin{thebibliography}{99}
\bibitem {Morse6}See for example, ref. 5, p. 314, where the high-temperature
expression for $U_{vib}$ is incorrect in Eq. (22-7).
\end{thebibliography}
\end{document}